\documentclass[%
 reprint,
aps,
nofootinbib]{revtex4-1}
\usepackage{scrextend}
\usepackage{graphicx}% Include figure files
\usepackage{dcolumn}% Align table columns on decimal point
\usepackage{bm}% bold math
\usepackage{float}
\usepackage[mathlines]{lineno}% Enable numbering of text and display math
\usepackage{footnote}
\usepackage{amsmath}
\usepackage{xfrac}
\usepackage{braket}
\usepackage{cancel}
\usepackage{xcolor}
\usepackage{amssymb}
\usepackage{slashed}
\usepackage[mathscr]{euscript}
\usepackage{lipsum}
\usepackage{mathtools}
\usepackage{cuted}
\usepackage[export]{adjustbox}

\usepackage{subcaption}
\usepackage[colorlinks=true, urlcolor=blue, linkcolor=red]{hyperref}

\usepackage{mathtools}

\newcommand{\arXiv}[2]{\href{http://arxiv.org/pdf/hep-th/#1}{{\tt #2/#1}}}

\newcommand{\arXivold}[2]{\href{http://arxiv.org/pdf/#1}{{\tt #2/#1}}}

\usepackage{blindtext} % just for demo
\begin{document}

\preprint{APS/123-QED}

\title{Aspects of Non-local QED and the Weak Gravity Conjecture}% Force line breaks with \\
\author{Fayez Abu-Ajamieh}
 %\altaffiliation[Also at ]{Physics Department, UC Davis.}%Lines break automatically or can be forced with \\
 \email{fayezajamieh@iisc.ac.in}
\affiliation{%
Centre for High Energy Physics, Indian Institute of Science, Bangalore 560012, India
}
\author{Nobuchika Okada}
\email{okadan@ua.edu}
\affiliation{
Department of Physics and Astronomy; 
University of Alabama; Tuscaloosa; Alabama 35487; USA}

\author{Sudhir K. Vempati}
\email{vempati@iisc.ac.in}
\affiliation{Centre for High Energy Physics, Indian Institute of Science, Bangalore 560012, India}


\begin{abstract}
We use the Weak Gravity Conjecture (WGC) to investigate the impact of charge dequantization arising from non-local QED on the scale of non-locality of neutrinos. We find this scale to be $\lesssim 87$ TeV, which could be probed in future colliders. We also investigate the electric force and potential in non-local QED and use them to reformulate the Weak Gravity Conjecture (WGC) in non-local QED.

%\begin{description}
%\end{description}
\end{abstract}
\pacs{Valid PACS appear here}% PACS, the Physics and Astronomy
\maketitle

%\tableofcontents

\section{Introduction}\label{sec1}
Although an agreed-upon quantum theory of gravity is still lacking, string theory is arguably the best candidate we have. It is known that not all candidate Quantum Field Theories (QFTs) can be UV-completed to include gravity, as suggested by string theory. As such, only those that can be are considered consistent QFTs.

The Swampland Conjecture program \cite{Vafa:2005ui,Ooguri:2006in} aims at qualifying such theories. According to the conjecture, QFTs that can be UV-completed to include gravity are said to lie in the landscape, whereas those that cannot are said to lie in the swampland.

One of the proposals within the swampland program is the WGC \cite{Arkani-Hamed:2006emk}. The conjecture states that for any $U(1)$ gauge group with charge $q$, there must exist a particle of mass $m$ such that 
\begin{equation}\label{eq:Gauge WGC}
m \leq \sqrt{2} q M_{\text{P}},
\end{equation}
where $M_{\text{P}} = 2.4 \times 10^{18}$ GeV is the reduced Planck scale. The conjecture is mostly based on the decay of black holes. For black holes to be able to decay and avoid the issue of  remnants \cite{Susskind:1995da}, a particle satisfying eq. (\ref{eq:Gauge WGC}) has to exist. Eq. (\ref{eq:Gauge WGC}) is called the electric WGC, and it can be shown that when applied to magnetic monopoles, it can be reformulated as follows: 
\begin{equation}\label{eq:Magnetic_WGC}
\Lambda \lesssim q M_{\text{P}}.
\end{equation}
where $\Lambda$ is the scale of New Physics (NP). Eq. (\ref{eq:Magnetic_WGC}) is called the magnetic WGC. It suggests that any $U(1)$ gauge theory has a natural cut-off below the Planck scale. The local WGC has many interesting implications, such as charge quantization and the masslessness of the photon \cite{Abu-Ajamieh:2024gaw, Abu-Ajamieh:2024xic}.

The WGC was formulated for local QFTs, and a natural question that might arise is how to extend this to non-local QFTs. At this point, a disclaimer is in order. We will assume that the WGC holds true as well in the non-local case, at least in the form discussed in the present work. The formation of black holes and their evaporation in generic non-local QFTs and the formulation of the WGC in that context is not the focus of this work.

In this paper, we try to answer this question for a class of non-local QFTs with infinite derivatives inspired by string theory.\footnote{Similar work was done in Lee-Wick QED \cite{Abu-Ajamieh:2024woy}.} In such theories, the kinetic term is modulated by the exponential of an entire function of the d'Alembertian operator. For example, a toy model with a real scalar can be written as
\begin{equation}\label{eq:NQFT}
\mathcal{L}_{\text{NL}} = -\frac{1}{2}\phi e^{\frac{\Box + m^{2}}{\Lambda_{\phi}^{2}}}(\Box + m^{2})\phi,
\end{equation}
where $\Lambda_{\phi}$ is the scale of non-locality, and it's easy to see that when $\Lambda_{\phi} \rightarrow \infty$, the local case is retrieved. As can be seen from eq. (\ref{eq:NQFT}), the non-local form factor improves the UV behavior of the theory, making it super-renormalizable.\footnote{It was argued in \cite{Abu-Ajamieh:2023syy} that some theories might still need renormalization through a special scheme in order to avoid unacceptably large corrections to certain observables.} 

The non-local version of QED, which will be used to formulate the non-local WGC, is given by \cite{Biswas:2014yia}
\begin{equation}\label{eq:NLQED}
\mathcal{L}_{\text{NL}} = -\frac{1}{4}F_{\mu\nu}e^{\frac{\Box}{\Lambda^{2}_{g}}}F^{\mu\nu} + \frac{1}{2}\Big[i\overline{\Psi}e^{-\frac{\nabla^{2}}{\Lambda_{f}^{2}}}(\slashed{\nabla}+m)\Psi + h.c. \Big].
\end{equation}
Non-local QED is both gauge-invariant and anomaly-free \cite{Abu-Ajamieh:2023roj}. As we shall show, the behavior of non-local QED is drastically different from that of local QED at high energy. This will have important implications for the WGC.

This paper is organized as follows: In Section \ref{sec2} we review charge dequantization in non-local QED and discuss its implications in light of the local WGC. In Section \ref{sec3}, we discuss the photon sector, and extract the electric force and potential in non-local QED. In Section \ref{sec4} we formulate the non-local version of the WGC. In Section \ref{sec5} we discuss the potential of setting bounds on the photon's scale of non-locality from Cavendish-type experiments, and finally we conclude in Section \ref{sec6}.

\section{Charge Dequantization in Non-local QED}\label{sec2}
It was shown in \cite{Capolupo:2022awe,Abu-Ajamieh:2023txh} that non-local QED leads inevitably to charge dequantization. This can be seen by calculating the charge renormalization form factor from the $\gamma\overline{f}f$ vertex at tree-level
\begin{equation}\label{eq:NL_Q_deq}
F_{1}(0) \equiv Q = (1+\frac{m^{2}_{f}}{\Lambda_{f}^{2}}) e^{\frac{m^{2}_{f}}{\Lambda_{f}^{2}}} \simeq 1 + \frac{2m^{2}_{f}}{\Lambda_{f}^{2}} +\cdots,
\end{equation}
and we can see that non-locality leads to the electric charge deviating from unity. The implications of such charge dequantization were analyzed in \cite{Abu-Ajamieh:2023txh} and here we review the main results. 

It was shown in \cite{Foot:1992ui} that the only way charge can be dequantizated in the Standard Model (SM) is as follows:
\begin{align}
Q^{i}_{\nu_{L}} & = Q^{i}_{\nu_{R}} = \epsilon, \label{eq:Q_deq1} \\
Q^{i}_{e_{L}} & = Q^{i}_{e_{R}} = -1+\epsilon, \label{eq:Q_deq2}\\
Q^{i}_{u_{L}} & = Q^{i}_{u_{R}} = \frac{2}{3}+\frac{1}{3}\epsilon, \label{eq:Q_deq3}\\
Q^{i}_{d_{L}} & = Q^{i}_{d_{R}} = -\frac{1}{3}+\frac{1}{3}\epsilon, \label{eq:Q_deq4}
\end{align}
where $i=1,2,3$, and these conditions arise for the requirement of anomaly cancellation. As we can see from eq. (\ref{eq:Q_deq1}), neutrinos will inevitably acquire an electric charge. 

As demonstrated in \cite{Babu:1989tq, Babu:1989ex}, if neutrinos are Majorana fermions, then they must be neutral ($\epsilon = 0)$ and the electric charge is quantized in the SM. This implies that non-local QED is inconsistent with Majorana fermions.

On the other hand, if neutrinos are Dirac fermions, then it is possible for them to acquire an electric charge. Comparing eq. (\ref{eq:NL_Q_deq}), with eqs. (\ref{eq:Q_deq1}) - (\ref{eq:Q_deq4}), it is not hard see that charge dequantization in non-local QED can only be accommodated if the scale of non-locality is flavor-dependent, with
\begin{equation}\label{eq:epsilon}
\epsilon = \frac{2m_{l}^{2}}{\Lambda_{l}^{2}} = \frac{4m_{u}^{2}}{\Lambda_{u}^{2}} = \frac{2m_{d}^{2}}{\Lambda_{d}^{2}}.
\end{equation}

Given that the strongest bound on the charge of neutrinos is $|\epsilon| < 10^{-21}$ \cite{Gallinaro:1966zza, Marinelli:1982dg, Marinelli:1983nd, Baumann:1988ue}, we obtain the following bounds on the scale of non-locality
\begin{align}
\Lambda_{e} \gtrsim 2.29 \times 10^{4} \hspace{2mm} \text{TeV}, \label{eq:bound1} \\
\Lambda_{\mu} \gtrsim 4.74 \times 10^{6} \hspace{2mm} \text{TeV},  \label{eq:bound2} \\
\Lambda_{\tau} \gtrsim 7.95 \times 10^{7} \hspace{2mm} \text{TeV},  \label{eq:bound3} \\
\Lambda_{u} \gtrsim 1.37 \times 10^{5} \hspace{2mm} \text{TeV},  \label{eq:bound4} \\
\Lambda_{d} \gtrsim 2.09 \times 10^{5} \hspace{2mm} \text{TeV},  \label{eq:bound5} \\
\Lambda_{s} \gtrsim 4.16 \times 10^{6} \hspace{2mm} \text{TeV},  \label{eq:bound6} \\
\Lambda_{c} \gtrsim 8.03 \times 10^{7} \hspace{2mm} \text{TeV},  \label{eq:bound7} \\
\Lambda_{b} \gtrsim 1.87 \times 10^{8} \hspace{2mm} \text{TeV},  \label{eq:bound8} \\
\Lambda_{t} \gtrsim 1.09 \times 10^{10} \hspace{2mm} \text{TeV}. \label{eq:bound9} 
\end{align}
Now, let's apply the (local) WGC to neutrinos. Given eqs. (\ref{eq:Gauge WGC}) and (\ref{eq:Q_deq1}), we have
\begin{equation}\label{eq:WGC_nu1}
m_{\nu_{i}} \leq \sqrt{2}\epsilon M_{\text{P}},
\end{equation}
and given the upper bound on $\epsilon$, we find $m_{\nu_{i}} \lesssim 3.44$ MeV. The PDG \cite{ParticleDataGroup:2020ssz} provides the following upper limits on the neutrino masses
\begin{align}
m_{\nu_{e}} & < 1.1 \hspace{2mm} \text{eV}, \label{eq:nu_e_mass}\\
m_{\nu_{\mu}} & < 0.19 \hspace{2mm} \text{MeV}, \label{eq:nu_e_mass}\\
m_{\nu_{\tau}} & < 18.9 \hspace{2mm} \text{MeV}. \label{eq:nu_e_mass}
\end{align}
and thus we can see that while $m_{\nu_{e}}$ and $m_{\nu_{\mu}}$ safely satisfy this bound, it is nonetheless possible that $m_{\nu_{\tau}}$ might not. This furnishes an experimental test for either non-local QED or the WGC. More specifically, if experiment finds that $m_{\nu_{\tau}} \gtrsim 3.44$ MeV, then this would either imply that neutrinos are neutral, which would subsequently exclude non-locality (at least in the fermion sector); or it would invalidate the WGC if neutrinos are found to have an electric charge.

If we plug eq. (\ref{eq:epsilon}) in eq. (\ref{eq:WGC_nu1}), we find that
\begin{equation}\label{eq:WGC_nu2}
\Lambda_{\nu_{i}} \leq \sqrt{2\sqrt{2}m_{\nu_{i}}M_{\text{P}}},
\end{equation}
and using the PDG's upper bounds on neutrino masses, we get the following bounds on the scale of non-locality
\begin{align}\label{eq:Lambda_nu_bounds}
\Lambda_{\nu_{\tau}} & \lesssim 3.6 \times 10^{5} \hspace{2mm} \text{TeV},\footnotemark[2] \\
\Lambda_{\nu_{\mu}} &\lesssim 3.6 \times 10^{4} \hspace{2mm} \text{TeV},\\
\Lambda_{\nu_{e}} &\lesssim 87 \hspace{2mm} \text{TeV}.
\end{align}
This implies that non-locality could potentially be probed in future colliders, such as the muon collider or the 100-TeV collider.
\footnotetext[2]{If we use the upper bound on $m_{\nu_{\tau}} \simeq 3.44$ MeV as suggested by eq. (\ref{eq:WGC_nu1}), the bound becomes $\Lambda_{\nu_{\tau}} \simeq 1.5 \times 10^{5} \hspace{2mm} \text{TeV}$.}

On the other hand, if we apply the magnetic WGC in eq. (\ref{eq:Magnetic_WGC}) with $q = 10^{-21}$, we find that $\Lambda_{\nu_{i}} \lesssim 2.43$ MeV! This is clearly excluded, which means that either neutrinos are neutral and thus non-locality is excluded, or that the WGC is invalid. However, we should point out that the magnetic WGC was challenged in \cite{Saraswat:2016eaz}, although it was argued that the electric form would continue to hold.

We choose to be conservative and allow the possibility that the argument in \cite{Saraswat:2016eaz} could be correct. Thus, we ignore the magnetic WGC and its implications

\section{The Photon Sector}\label{sec3}
\begin{figure}[!t] 
\centering
\includegraphics[width=0.45\textwidth]{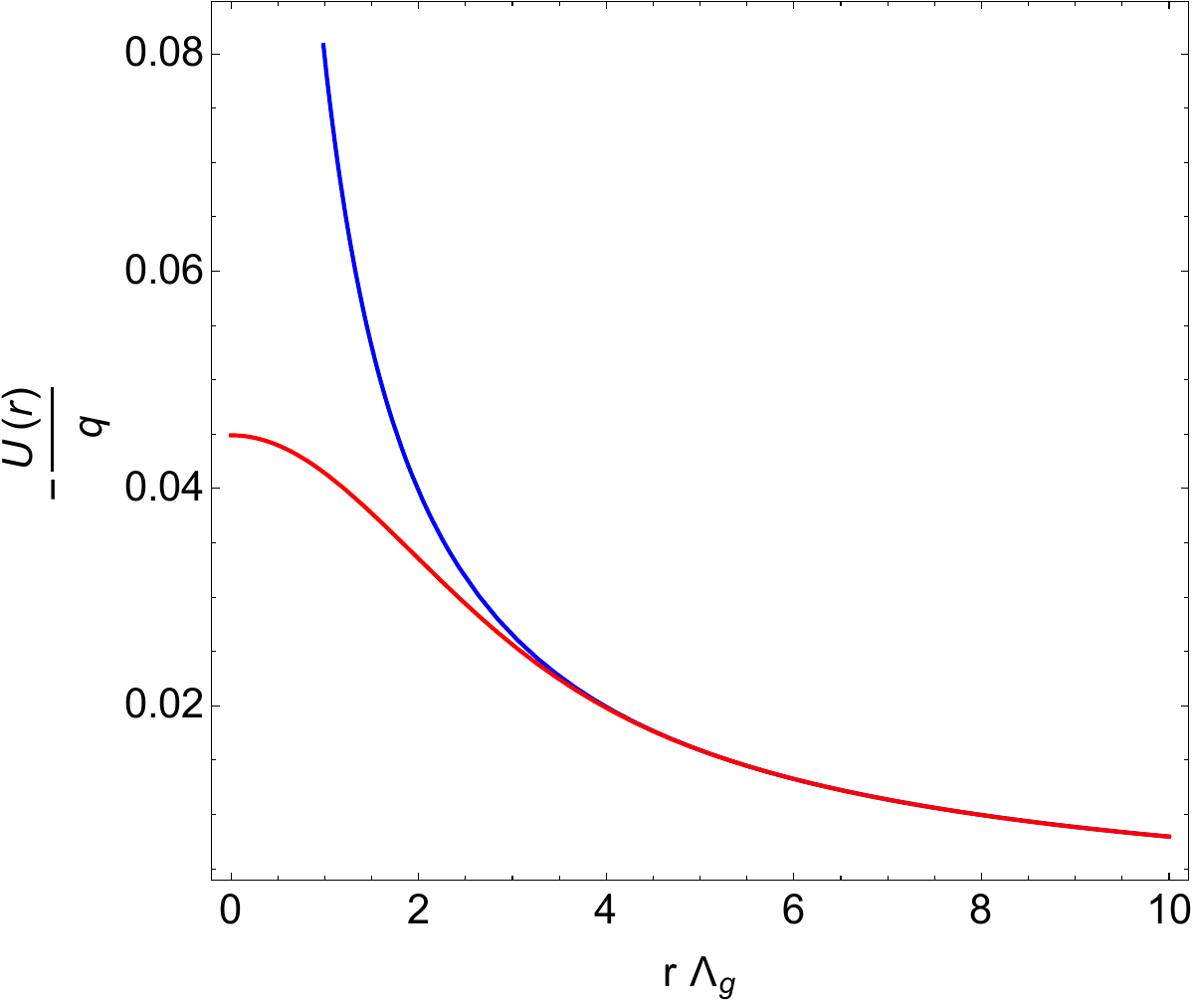}
\caption{The non-local (red) vs. local (blue) electric potentials.}
\label{fig1}
\end{figure}

Turning our attention to the photon sector, we find from eq. (\ref{eq:NLQED}) that the non-local photon propagator is given by
\begin{equation}\label{eq:PhotonProp}
\Pi_{g} = \frac{-i g_{\mu\nu}}{k^{2}} e^{\frac{k^{2}}{\Lambda_{g}^{2}}},
\end{equation}
and in general $\Lambda_{g}$ is assumed to be independent of $\Lambda_{f}$. The non-local electric potential corresponding to eq. (\ref{eq:PhotonProp}) is given by
\begin{align}\label{eq:NL_potential}
U(r)  & = \lim_{k_{0}\rightarrow 0} \int \frac{d^{3}\vec{k}}{(2\pi)^{3}} e^{i \vec{k}.\vec{r}}\frac{q\exp{\Big(\frac{k_{0}^{2} - |\vec{k}|^{2}}{\Lambda_{g}^{2}}}\Big)}{k_{0}^{2} - |\vec{k}|^{2}}, \nonumber\\
& = -\frac{q}{4\pi r} \text{erf}\Big( \frac{1}{2} r \Lambda_{g}\Big),\nonumber\\
& = U_{\text{Coul}}(r) \times \text{erf}\Big( \frac{1}{2} r \Lambda_{g}\Big),
\end{align}
where $\text{erf}(x)$ is the error function and $U_{\text{Col}}$ is the (local) Coulomb potential, and we see from eq. (\ref{eq:NL_potential}) that non-locality modulates the Coulomb potential. In Figure \ref{fig1}, we plot the local and non-local electric potentials. As can be seen from the plot, the non-local potential approaches the local one when $ r \Lambda_{g} \gg 1$, which corresponds to the low energy limit, however, in the regime $ r \Lambda_{g} \ll 1$ corresponding to the high energy limit, the non-local potential deviates drastically from the local case: More specifically, while the Coulomb potential blows up in the limit $r \rightarrow 0$, the non-local case remains finite. In fact, we find that 
\begin{equation}\label{eq:NL_potential_limit}
\lim_{r \rightarrow 0} U_{\text{NL}}(r) = -\frac{q\Lambda_{g}}{4\pi^{3/2}}.
\end{equation}

This behavior can be understood by inspecting the propagator in eq. (\ref{eq:NLQED}) and eq. (\ref{eq:PhotonProp}): When $\Lambda_{g} \rightarrow \infty$, the form factor becomes unity and the local case is retrieved (which is the case for all such non-local QFTs by construction). On the other hand, when $|\vec{k}|^{2} \gg \Lambda_{g}^{2}$, then $\Pi_{g} \sim e^{-k^{2}/\Lambda_{g}^{2}} \rightarrow 0$ and one can see how non-locality removes UV divergences.

It is now a simple exercise to find the non-local electric force corresponding to eq. (\ref{eq:NL_potential}) by taking the derivative with respect to $r$
\begin{equation}\label{eq:NL_force}
F_{\text{NL}} (r)  = -\frac{q}{4\pi r^{2}} \text{erf} \Big(\frac{1}{2}r \Lambda_{g} \Big) + \frac{q \Lambda_{g}}{4\pi^{3/2}r} e^{-\frac{1}{4}r^{2}\Lambda_{g}^{2}}.
\end{equation}

\begin{figure}[!t] 
\centering
\includegraphics[width=0.45\textwidth]{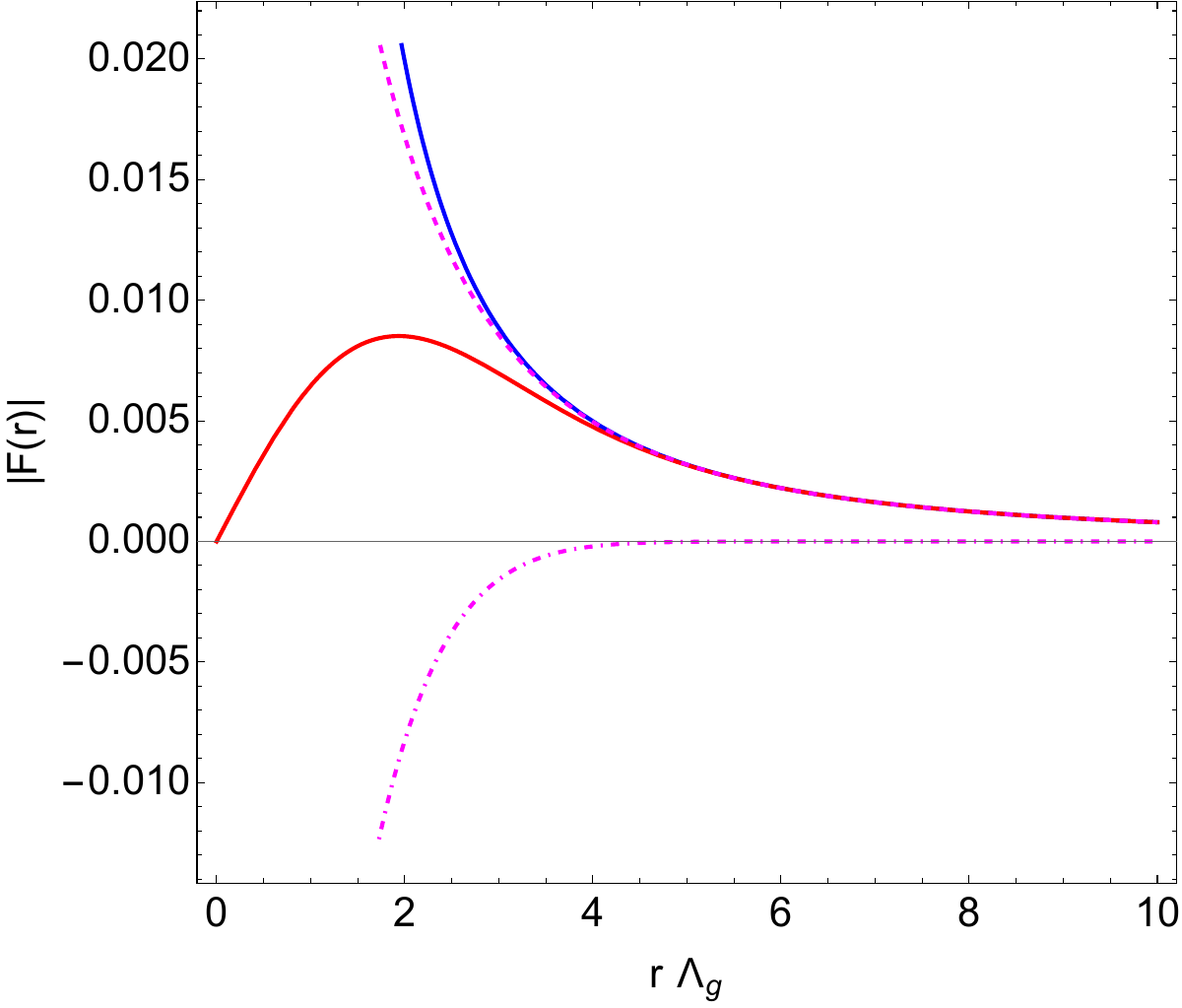}
\caption{The non-local (red) vs. local (blue) electric force. The dashed line corresponds to the first term in eq. (\ref{eq:NL_force}), whereas the dot-dashed line corresponds to the second term in eq. (\ref{eq:NL_force}).}
\label{fig2}
\end{figure}

We plot the non-local force in Figure \ref{fig2}, together with the Coulomb force. Here too, we find a similar behavior to the potential as expected: The non-local force approaches the local one in the low energy regime, and deviates drastically from it in the high energy regime. We see that while the Coulomb force blows up when $r \rightarrow 0$, the non-local force vanishes in that limit. An interesting thing to observe, is that while the two contributions to the non-local force do blow up in the limit $r \rightarrow 0$, their sum nonetheless remains finite.

This deviation between the local and non-local force/ potential suggests that non-locality can be probed through fifth force searches and Cavendish-type experiments. We shall investigate the possibility of probing the scale of non-locality $\Lambda_{g}$ using the results of Cavendish-type experiments in Section \ref{sec5}, but before we do so, let's first reformulate the WGC in non-local QED.

\section{The Non-local WGC}\label{sec4}
\begin{figure}[!t] 
\centering
\includegraphics[width=0.45\textwidth]{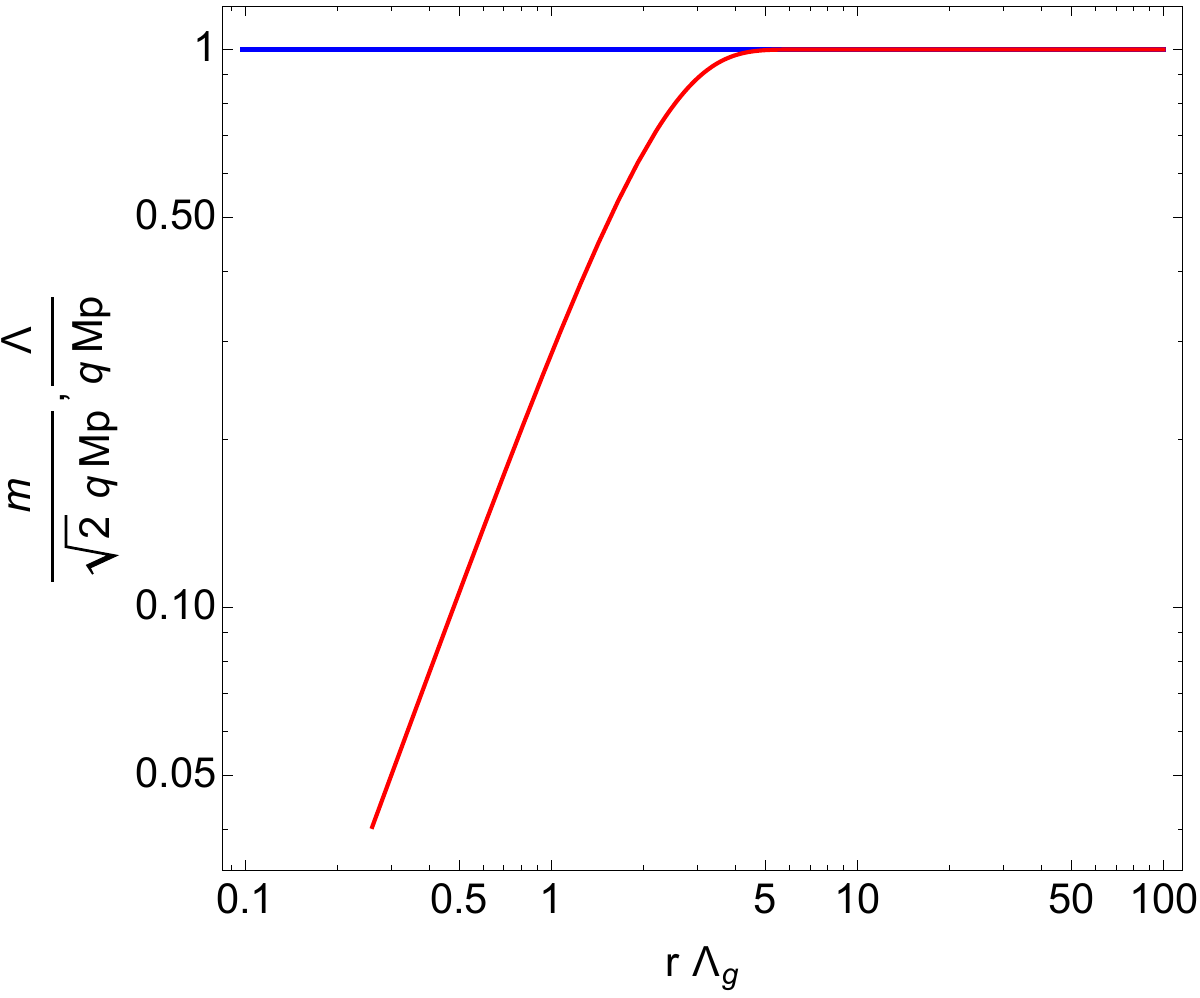}
\caption{The non-local WGC (red) and the local WGC (blue), normalized such that the latter $ =1$.}
\label{fig3}
\end{figure}

The physical content of the local WGC is that the gravitational force is weaker than any $U(1)$ force, i.e.
\begin{equation}\label{eq:WGC_content}
F_{\text{Grav}} = \frac{m^{2}}{8\pi M_{\text{P}}^{2} r^{2}} \leq F_{U(1)} = \frac{q^{2}}{4\pi r^{2}},
\end{equation}
which immediately leads to eq. (\ref{eq:Gauge WGC}). To obtain the WGC in the non-local case, we simply replace the local force with the non-local one from eq. (\ref{eq:NL_force}). We obtain
\begin{equation}\label{eq:NL_WGC}
m \leq \sqrt{2} q M_{\text{p}}\Big[ \text{erf} \Big( \frac{1}{2}r \Lambda_{g}\Big) - \frac{r \Lambda_{g}}{\sqrt{\pi}} e^{-\frac{1}{4}\Lambda_{g}^{2} r^{2}} \Big]^{1/2},
\end{equation}
and we can see that the non-local WGC differs from the local one by the term in the square brackets. We plot the local and non-local WGCs in Figure \ref{fig3}, and we normalize them such that the former is set to unity. As usual, we find that both versions agree in the low energy limit ($r \Lambda_{g} \gg 1$), and we see that in the limit $r \rightarrow \infty$, the term in the square brackets tends to 1. On the other hand, in the high energy limit ($r \Lambda_{g} \ll 1$), the non-local factor becomes smaller, which means that the non-local case becomes more stringent than the local case. This is most apparent when we inspect the magnetic WGC in the non-local case
\begin{equation}\label{eq:NL_mag_WGC}
\Lambda \lesssim  q M_{\text{p}}\Big[ \text{erf} \Big( \frac{1}{2}r \Lambda_{g}\Big) - \frac{r \Lambda_{g}}{\sqrt{\pi}} e^{-\frac{1}{4}\Lambda_{g}^{2} r^{2}} \Big]^{1/2},
\end{equation}
where $\Lambda$ is the scale of NP, not to be confused with the scale of non-locality. It is easy to see that the scale of NP will always be \textit{lower} than the local case.

This behavior is plausible, as non-locality reduces the UV growth in physical quantities, which means that the scale of NP must also be reduced compared to the local case.

Another interesting behavior of the non-local WGC is that now it depends on $r$, i.e., it depends on the energy scale itself. This introduces ambiguity with regards to which energy scale should be used. This should probably be the length/ energy scale of the experiment at hand. In general, we expect that $r$ could range between the Planck scale $\sim M_{\text{P}}^{-1}$ and the Hubble radius $\sim 10^{-33}$ eV.

\section{Probing $\Lambda_{g}$ through Cavendish-type Experiments}\label{sec5}
As we saw in Section \ref{sec3}, the non-local electric potential deviates from the local one, especially at high energy. Therefore, we can use Cavendish-type experiments to set limits on the photon's scale of non-locality $\Lambda_{g}$. 

In general, deviation from the Coulomb potential is parametrized in two ways:
\begin{itemize}
\item Assuming $U(r) = \frac{q}{4\pi r^{1+\delta}}$ and setting limits on $\delta$, or
\item Assuming $U(r) = \frac{q}{4\pi r}e^{-\mu r}$ and setting limits on the photon mass $\mu$.
\end{itemize}

The most stringent bounds from Cavendish-type experiments were obtained in \cite{Williams:1971ms} (see also \cite{ Fulcher:1986}), with the bounds on the photon mass given by
\begin{equation}\label{eq:PhotonMass}
\mu^{2} = (1.04 \pm 1.2) \times 10^{-19} \hspace{2mm} \text{cm}^{-2}.
\end{equation}

Comparing eq. (\ref{eq:NL_potential}) with the second parametrization, and requiring that the deviation due to non-locality be within experimental range, we obtain the following condition
\begin{equation}\label{eq:Cavendish}
\frac{1}{r} \log{\Big[ \text{erf}\Big( \frac{1}{2}r\Lambda_{g}\Big)\Big]} + \mu_{\text{max}} \leq 0.
\end{equation}

The experiment in \cite{Williams:1971ms} used 4 concentric spheres of radii $r_{1} = 49.54$ cm, $r_{2} = 78$ cm, $r_{3} = 78.53$ cm and $r_{4} = 105.41$ cm. Thus, we can set $r$ to one of these radii. However, the bound obtained is very weak ($\sim O(10^{-15})$ GeV). Thus, Cavendish-type experiments are not sensitive enough to probe non-locality.

Before we conclude, we point out that in Section \ref{sec2}, bounds were derived based on the \textit{local} WGC, and given that the non-local WGC is always more stringent than the local one. Thus it is worthwhile trying to reevaluate these limits using the non-local WGC. 

Inspecting eqs. (\ref{eq:NL_WGC}) and (\ref{eq:NL_mag_WGC}), we see that one needs both $r$ and $\Lambda_{g}$. Bounds on the charge of neutrinos are obtained from the bounds on matter neutrality. The upper limit on the charge of neutrinos was obtained in \cite{Gallinaro:1966zza, Marinelli:1982dg, Marinelli:1983nd, Baumann:1988ue}. There, a magnetic field was used to create a magnetic potential valley that was used to levitate graphite grains. A horizontal electric field was then applied between two plates 1.5 mm apart, causing the particles to be displaced from the equilibrium position. The deviation between the two positions was then used to set bounds on the fractional charge carried by the graphite grain, which could be translated into bounds on the charge of neutrinos.

Therefore, it is plausible to set $r = 1.5$ mm when using $|\epsilon| < 10^{-21}$. On the other hand, choosing a value for $\Lambda_{g}$ is more problematic, as in general one can place an upper limit on $\Lambda_{g}$, and increasing it will only loosen the bound until it eventually matches the limit from the local case. Therefore, one cannot use the limit in eq. (\ref{eq:NL_mag_WGC}) without knowing $\Lambda_{g}$. Thus, to be conservative, we confine ourselves to the limits obtained from the local WGC. 

\section{Conclusions}\label{sec6}
In this paper, we studied certain aspects of non-local QED, including the WGC. We used charge dequatization arising from non-locality in the (local) WGC to set an upper bound on the masses of neutrinos $\simeq 3.44$ MeV, which is satisfied by $\nu_{e}$ and $\nu_{\mu}$, but could be violated by the mass of $\nu_{\tau}$, furnishing a test for either non-locality or the WGC. We also set bounds on the scale on non-locality for neutrinos and showed that for $\nu_{e}$, it should be $\lesssim 87$ TeV, which could be probed in future colliders.

We also extracted the electric force and potential corresponding to the non-local QED, and found that while the Coulomb potential and force blow up as $r \rightarrow 0$, their non-local counterparts remain finite. We used the non-local force to formulate the non-local version of the WGC and showed that it implies a smaller mass and a lower cutoff scale compared to the local WGC. We found that Cavendish-type experiments are not sensitive enough to probe the scale of non-locality.

\section*{Acknowledgments}
The work of NO is supported in part by the United States Department of Energy (DC-SC 0012447 and DC-SC 0023713). SKV is supported by SERB, DST, Govt. of India Grants MTR/2022/000255 , “Theoretical aspects of some physics beyond standard models”, CRG/2021/007170 “Tiny Effects from Heavy New Physics “and IoE funds from IISC. 
\appendix

\end{document}